\documentclass[conference,compsoc]{IEEEtran}

\IEEEoverridecommandlockouts
\usepackage{cite}
\usepackage{amsmath,amssymb,amsfonts}
\usepackage{algorithmic}
\usepackage{graphicx}
\usepackage{textcomp}
\usepackage{xcolor}
\usepackage[hidelinks]{hyperref}
\usepackage{xurl}
\usepackage{booktabs}
\usepackage{tabularx}
\def\BibTeX{{\rm B\kern-.05em{\sc i\kern-.025em b}\kern-.08em
    T\kern-.1667em\lower.7ex\hbox{E}\kern-.125emX}}

\begin{document}
\title{Applying Public Health Systematic Approaches to Cybersecurity: The Economics of Collective Defense}

\author{
\IEEEauthorblockN{Josiah Dykstra}
\IEEEauthorblockA{\textit{Designer Security, LLC} \\
Ellicott City, MD USA \\
josiah@designersecurity.com}
\and

\IEEEauthorblockN{William Yurcik*}
\thanks{* Disclaimer: The views presented herein do not represent the views of the Federal Government.}
\IEEEauthorblockA{\textit{Centers for Medicare \& Medicaid Services (CMS)} \\
Baltimore, MD USA \\
William.Yurcik@cms.hhs.gov}
}

\maketitle

\begin{abstract}
The U.S. public health system increased life expectancy by more than 30 years since 1900 through systematic data collection, evidence-based intervention, and coordinated response. This paper examines whether cybersecurity can benefit from similar organizational principles. We find that both domains exhibit public good characteristics: security improvements create positive externalities that individual actors cannot fully capture, leading to systematic market failure and underinvestment. Current cybersecurity lacks fundamental infrastructure including standardized population definitions, reliable outcome measurements, understanding of transmission mechanisms, and coordinated intervention testing. Drawing on public health's transformation from fragmented local responses to coordinated evidence-based discipline, we propose a national Cyber Public Health System for systematic data collection, standardized measurement, and coordinated response. We argue government coordination is economically necessary rather than merely beneficial, and outline specific federal roles in establishing standards, funding research, coordinating response, and addressing information asymmetries that markets cannot resolve.

\end{abstract}


\section{Introduction}
\label{section:Introduction}
In 1854, a cholera outbreak devastated London. While the medical establishment attributed disease spread to ``miasma'' or bad air, physician John Snow suspected contaminated water as the culprit~\cite{Johnsnowpaper}. His investigation was elegantly simple: he mapped each cholera death as a dot and marked the locations of public water pumps. The resulting visualization revealed a striking cluster of deaths around a single pump on Broad Street~\cite{Johnsnowpaper}. Local officials removed the pump handle, and the outbreak ended~\cite{topographyofcholera}. Snow's success came not from advanced technology, but from systematic data collection, spatial analysis, and evidence-based intervention---the foundational principles of epidemiology.

Cybersecurity today faces a similar challenge. Despite decades of research and billions in spending, the cybersecurity community lacks the systematic infrastructure to answer fundamental questions: How many data breaches actually occurred last year? Which defensive measures demonstrably reduce risk? Are we safer this year than last? Like 19th-century physicians debating miasma theory while lacking the framework to test their hypotheses, defenders deploy cybersecurity measures without rigorous evidence of their effectiveness at a population level. Individual organizations respond to incidents reactively, share threat intelligence within limited circles, and struggle to translate localized successes into broad improvements.

The contrast with modern public health is stark. The U.S. public health system has increased life expectancy by more than 30 years since 1900, with 25 of those years directly attributable to public health advances~\cite{cdcpaper}. This remarkable achievement stems from a coordinated national infrastructure for systematic data collection, standardized measurement, evidence-based interventions, and continuous monitoring. Public health practitioners can quantify disease prevalence, track transmission vectors, measure intervention effectiveness, and allocate resources based on empirical evidence. When COVID-19 emerged, this infrastructure enabled rapid detection, coordinated response, and continuous adaptation based on emerging data.

Cybersecurity has no equivalent infrastructure. The cybersecurity community is not reliably measuring the ``cyber health'' of the nation's digital infrastructure. The field lacks lack standardized definitions of fundamental concepts like ``population'' in cyberspace. Practitioners have limited ability to track how cyber threats spread or to conduct the equivalent of public health's environmental surveillance. Most critically, the community operates without a coordinated system for testing whether interventions actually work at scale. Each organization fights its own battles, learning its own lessons, often repeating mistakes others have already made.

This paper argues for a fundamental reorganization of cybersecurity around a national Cyber Public Health System as a coordinated federal infrastructure that would bring public health's systematic approach to cyber defense. Drawing on the principles that transformed public health from sporadic local efforts into an evidence-based discipline, we propose a framework for systematic data collection, standardized measurement, coordinated response, and continuous improvement in cybersecurity.

The parallel between public health and cybersecurity is imperfect but instructive. A critical difference actually strengthens the case for this approach: unlike human health, compromised computer systems can be forensically investigated, wiped clean, and rebooted. This ``reversibility'' provides cybersecurity with an opportunity public health lacks: the ability to conduct controlled interventions, measure their effects, and iterate rapidly. Rather than making cyber threats less serious, this characteristic makes cybersecurity an ideal domain for evidence-based practice.

The remainder of this paper is organized as follows. Section~\ref{section:Motivation} describes the motivation for this paper. Section~\ref{section:Previous Work} highlights related work on this topic. 
Section~\ref{section:Known Unknowns in Cybersecurity} scopes the challenge faced by this effort by describing the known unknowns in cybersecurity and Section~\ref{section:Measuring Progress} describes the measurement challenges. 
Section~\ref{section:Argument for Government Involvement} argues that cybersecurity exhibits public good characteristics necessitating Federal Government engagement. We end with a short summary in Section~\ref{section:Summary} highlighting how lessons from public health can bridge the gap between cybersecurity and policymakers, fostering a more holistic and effective defense against emerging cyber threats.

\section{Motivation}
\label{section:Motivation}

\begin{figure} [ht]
    \centering
    \includegraphics[width=1.00\linewidth]{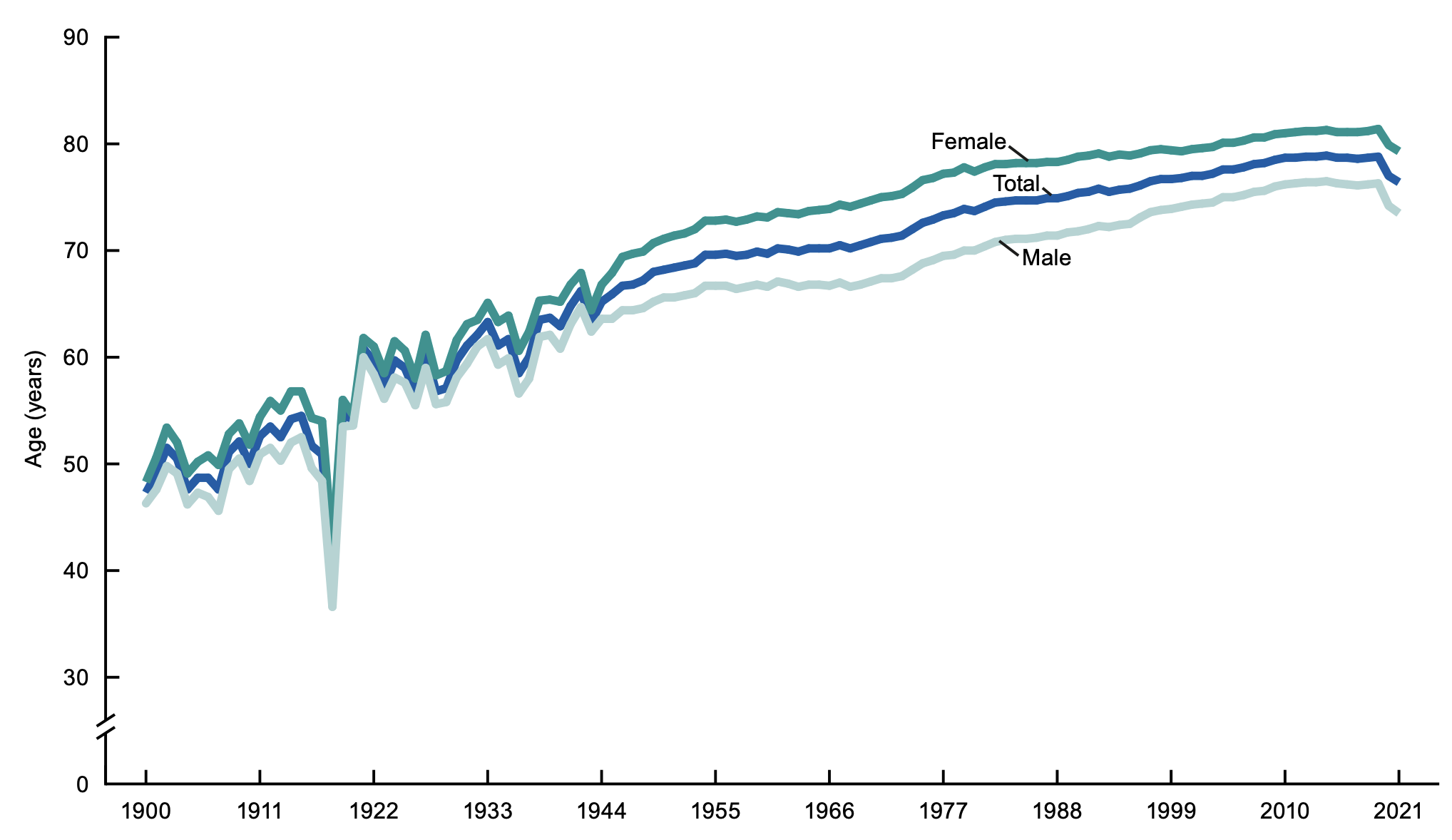}
    \caption{U.S. Life Expectancy 1900-2021~\cite{cdc2023life}. }
    \label{fig:US_Life_Expectancy}
\end{figure}

The U.S. public health system has achieved remarkable success in addressing infectious and chronic diseases, resulting in a significant increase in life expectancy. As shown in Figure~\ref{fig:US_Life_Expectancy}, the average life expectancy in the U.S. has increased by more than 30 years since 1900, with about 25 of those years directly attributable to public health advances~\cite{cdcpaper}.  The U.S. life expectancy at birth has roughly doubled since 1850, from about 39.4 years to nearly 80 years in recent data~\cite{cdcpaper}. 

This transformation required fundamental changes in how society organized to protect population health. Throughout recorded history, epidemics such as plagues, cholera, and smallpox evoked sporadic public efforts to protect citizens, but these remained fragmented and reactive. Modern public health emerged in the 18th and 19th centuries as systematic understanding replaced superstition. Early organized efforts include the 1798 establishment of the U.S. Marine Hospital Service as a federal response to care for sick sailors---an early recognition that health threats transcend individual or local capacity to respond. The foundation of modern public health was significantly shaped by the Public Health Act of 1848 in Britain, which created the first national framework for sanitation, and the subsequent development of scientific understanding of germ theory and disease prevention~\cite{publichealthhistory}.

The transition from reactive, localized responses to a coordinated national system enabled specific, measurable improvements. Childhood vaccines and improved sanitation controlled infectious diseases that once killed millions. Strategic vaccination campaigns virtually eliminated diseases that were common in the U.S., including diphtheria, tetanus, polio, smallpox, measles, mumps, rubella, and meningitis~\cite{cdcpaper2}. In 1900, approximately one in five children died before their fifth birthday; by 2024, this rate had fallen below 1 in 200~\cite{owid-vaccines-children-saved}. Maternal and infant mortality declined dramatically. Food and workplace safety improved through systematic regulation and monitoring.

The dramatic decline in infectious disease mortality through the 20th century demonstrates what coordinated public health achieves. However, history also reveals the consequences of complacency. Success in reducing infectious disease mortality during the first three quarters of the 20th century led to decreased investment in treatment and control of infectious microbes. This complacency contributed to the emergence of AIDS and the resurgence of tuberculosis (including multidrug-resistant strains), resulting in increased infectious disease mortality during the 1980s and early 1990s~\cite{publichealthhistory2}. The dips visible in Figure~\ref{fig:US_Life_Expectancy}---first from the Spanish Flu pandemic and later from COVID-19—underscore that as long as pathogens can evolve, new threats will emerge. Continuous vigilance, systematic monitoring, and coordinated response capability remain essential.

The public health system accomplished these results through specific organizational principles and practices:

\textbf{Systematic Data Collection}: Standardized definitions enable consistent measurement. Public health practitioners know how many cases of measles occurred last year because reporting follows uniform protocols. They know which populations are at elevated risk because demographic data is systematically collected and analyzed.

\textbf{Evidence-Based Interventions}: Public health tests whether interventions work. Vaccine efficacy is measured through rigorous trials. Sanitation improvements are evaluated for their impact on disease transmission. Resources flow to interventions with demonstrated effectiveness.

\textbf{Coordinated Response}: Individual hospitals cannot stop epidemics alone. The public health infrastructure enables information sharing, resource allocation, and coordinated action at the local, state, and federal levels. The U.S. Centers for Disease Control and Prevention (CDC) serves as a central coordinating body that synthesizes information and provides guidance based on the best available evidence.

\textbf{Continuous Monitoring and Adaptation}: Disease surveillance systems detect emerging threats early. When new diseases appear or known diseases develop resistance, the system adapts. Public health does not simply deploy interventions and hope for the best; it continuously measures outcomes and adjusts strategies accordingly.

\textbf{Population-Level Thinking}: Public health addresses threats that transcend individual action. Herd immunity protects those who cannot be vaccinated. Clean water systems protect entire communities. The framework recognizes that some challenges require collective action and public coordination.

\textbf{Recognition as a Public Good:} Fundamentally, public health operates on the recognition that population health is a \textit{public good} requiring collective investment and coordination~\cite{chen1999health}. When individuals receive vaccinations, they create positive externalities in the form of herd immunity that protects others who cannot be vaccinated. When sanitation systems are improved, entire communities benefit regardless of individual contributions to the cost. These characteristics mean that markets alone systematically under-provide population health interventions, as private actors cannot fully capture the social value they create. This economic reality explains why public health requires government coordination and funding rather than relying solely on private healthcare markets. The same logic may apply to cybersecurity: if improved digital security creates positive externalities that benefit the broader ecosystem, then market-based approaches alone will systematically under-invest in cyber defenses.

The stakes are substantial and growing. Cybersecurity is critical infrastructure for modern society. Disruptions threaten not just individual organizations but economic stability, national security, and public safety. Large-scale attacks can disable essential services: energy grids, water systems, transportation networks, financial systems, healthcare delivery, and emergency services. The SolarWinds compromise demonstrated how adversaries can penetrate thousands of organizations through a single supply chain vulnerability. The Colonial Pipeline ransomware attack showed how a single incident can cause widespread fuel shortages and economic disruption. The healthcare sector experiences regular attacks that compromise patient data and disrupt medical services.

Yet despite these escalating threats and growing dependence on digital infrastructure, cybersecurity defenders continue to fight cyber threats without the systematic infrastructure that public health developed over the past century. The cybersecurity community lacks standardized metrics for measuring cyber health. We cannot reliably track how threats spread through digital ecosystems. We deploy defensive measures without rigorous evidence of their effectiveness at scale. We respond to incidents post-compromise rather than detecting and containing threats early through systematic surveillance.

The motivation for this paper is straightforward: if a coordinated national public health system successfully increased human life expectancy by 30 years, could a similar approach to organizing cybersecurity efforts significantly improve our ability to protect critical digital infrastructure? Given public health's proven model for addressing evolving threats through systematic data collection, evidence-based intervention, and coordinated response, now is the appropriate time to advocate for applying these principles to protect the ``digital health'' of our nation's cyber infrastructure.

\section{Previous Work}
\label{section:Previous Work}

The application of public health principles to cybersecurity has emerged as a growing area of research over the past two decades, driven by recognition that both domains face analogous challenges in protecting interconnected populations from evolving threats.

Several organizations have actively advocated for cyber public health approaches. The CyberGreen Institute has been a leading voice, arguing that adopting a public health perspective embracing data-driven investigation, population thinking, and preventative approaches would transform cybersecurity practice~\cite{cybergreen2022public}. In January 2024, Google and CyberGreen hosted a workshop with 35 researchers to brainstorm comparisons between problem-solving in cybersecurity and public health~\cite{Shostak_2024}. Three high-level points of consensus emerged: (1) cyber public health is a field unto itself, (2) specific tangible actions can be taken immediately such as collecting appropriate data, and (3) community externalities of cybersecurity can help at large scale, paralleling public health externalities.

This paper contributes to ongoing policy discussions about cybersecurity organization. Our work builds on a research paper for the 2024-25 Computing Research Association (CRA) Quadrennial Papers, which identified cybersecurity informed by public health principles as having potential to address national priorities~\cite{shostack2025lessons}. That work, along with the present paper, argues that national institutions and international frameworks drawing on public health lessons could help create a safer, more secure Internet.

Several studies have examined specific parallels between the domains of cybersecuritiy and public health. Research has explored similar language, analogies, and even myths shared between public health and cybersecurity~\cite{dykstra2024handling, dykstra2023position, spafmythbook}. 

The foundation for applying economic analysis to cybersecurity was established by Anderson, who demonstrated that many security failures stem from misaligned economic incentives rather than technical limitations~\cite{anderson2001information}. His seminal work showed that when parties who could fix security problems do not bear the costs of failure, systems remain vulnerable. Anderson and Moore later expanded this analysis, demonstrating how market failures, externalities, and information asymmetries systematically undermine security~\cite{anderson2006economics}. Varian identified that system reliability behaves as a \textit{public good} with non-excludable and non-rivalrous benefits, establishing the theoretical foundation for treating cybersecurity improvements as generating positive externalities~\cite{varian2004system}. Researchers have also applied the concept of public goods to the Internet and cybersecurity~\cite{naghizadeh2016provision}.

\section{Known Unknowns in Cybersecurity}
\label{section:Known Unknowns in Cybersecurity}
Public health's systematic approach rests on three foundational concepts that enable evidence-based practice: clearly defined \textit{populations} at risk, reliably measured \textit{outcomes} that track health status, and well-understood \textit{transmission mechanisms} that explain how diseases spread. Each concept supports the others—understanding which populations face elevated risk enables targeted interventions, measuring outcomes shows whether interventions work, and knowing transmission mechanisms guides both prevention and response strategies. This integrated framework transformed public health from reactive crisis management to proactive population protection.

Cybersecurity lacks clarity on all three dimensions. Despite decades of research and billions in investment, we cannot answer fundamental questions that public health routinely addresses. The ambiguity begins with the most basic question: what constitutes a ``population'' in cybersecurity? Is it individual users, devices, accounts, networks, organizations, or security incidents themselves? Without standardized definitions of what we are protecting, meaningful measurement becomes impossible. Public health can state with confidence how many people contracted measles last year because ``population'' has clear, agreed-upon definitions. Cybersecurity has no equivalent consensus.

The outcomes we seek to measure remain similarly ill-defined. In public health, where the goal is sustained life and well-being, measurement is straightforward: life expectancy, quality-adjusted life years, disease incidence and prevalence. What are the analogous measures in cybersecurity? Time systems remain uncompromised? Number of successful intrusions prevented? Economic value protected? The cybersecurity community does not know how many data breaches occurred last year. We have numbers, but they are not  accurate enough because different organizations define ``breach'' differently, reporting is inconsistent, and many incidents go undetected or unreported. The result is that we cannot know how much money should be budgeted to prevent future breaches. Any number is a bad number, but accuracy toward the order of magnitude is important for resource allocation.

The knowledge gaps extend beyond measurement to basic infrastructure information. The medical community knows how many doctors practice in the U.S. because medicine is a licensed profession with mandatory registration. In cybersecurity, we do not know how many cybersecurity professionals work in the U.S. because the field lacks professional licensing. We know how many hospitals exist and how many inpatient beds they operate. We even know how many specialty care hospitals have trained staff and equipment to handle specific, high-acuity health conditions. This asset inventory enables public health to make informed decisions about resource distribution during crises. Similar information for cybersecurity would be valuable, such as knowing which organizations have mature security operations, incident response capabilities, and threat intelligence functions, but we lack both the data and access to analyze it systematically.

Perhaps most critically, we lack understanding of how cyber threats propagate. Public health identifies disease transmission mechanisms with precision, like whether pathogens spread through vectors like rodent fleas (plague), airborne transmission (tuberculosis), respiratory droplets (influenza), or direct contact (Ebola). Knowing the transmission mechanism enables targeted monitoring through sampling and environmental surveillance, informs public education about risk-reducing behaviors, and guides intervention strategies. If a disease spreads through contaminated water, improve water systems; if airborne, improve ventilation and use masks; if vector-borne, control the vector population.

Cybersecurity struggles with analogous understanding. Malware transmission mechanisms often remain opaque. We observe that systems become infected but struggle to trace precisely how threats spread across interconnected networks. Is it through unpatched vulnerabilities, social engineering, supply chain compromise, insider threats, or some combination? Without systematic environmental surveillance equivalent to public health's pathogen monitoring, we cannot identify emerging threats early or track their spread reliably. Individual organizations see attacks against their own systems but lack visibility into broader patterns. Attackers, conversely, observe their techniques across many targets and learn what works—creating an information asymmetry that favors offense.

If we do not have a benchmark standard of reference for measuring populations, outcomes, and transmission mechanisms, then we cannot determine whether we are safer today than yesterday. Public health can state with confidence that infectious disease mortality declined through the 20th century because standardized measurement enables comparison over time. Cybersecurity cannot make equivalent claims about improving ``cyber health'' because we lack the foundational infrastructure for measurement.

These gaps persist not because of insufficient technical capability or research effort, but because cybersecurity lacks the coordinated institutional infrastructure that public health developed over more than a century. The question is not whether we have the technical means to collect data, define populations, and track threats--we do. The question is whether we can organize collective action to build the systems for doing so systematically. This requires understanding why such infrastructure does not emerge spontaneously from market forces alone.

\section{Measuring Progress}
\label{section:Measuring Progress}

It has been said that you cannot manage something that is not being measured. If public health has made major improvements as measured in increased lifespan, what would making measured improvements in cybersecurity look like? What would we see?

Public health tracks population-level outcomes through standardized reporting systems. Disease incidence and prevalence, mortality rates, vaccination coverage, and quality-adjusted life years provide clear signals of population health status. These metrics enable comparison across regions and time periods, allowing practitioners to identify what works. A comparable approach to cybersecurity would track observable outcomes at scale: statistical trends in successful cyber attacks, time from compromise to detection, prevalence of known vulnerabilities in deployed systems, and adoption rates of effective security controls. Such metrics would be equivalent to the periodic public sharing of public health statistics, except focused on the ``cyber health'' of digital infrastructure.

Public health measures not just outcomes but the effectiveness of specific interventions. Vaccine efficacy is determined through rigorous trials before widespread deployment. Sanitation improvements are evaluated for their impact on disease transmission. Resources flow to interventions with demonstrated effectiveness. Cybersecurity could adopt similar rigor. Some security measures already have measurable effects that could be systematically tracked. A shorter patching cadence has been documented to correlate with reduced risk, as it narrows the window during which systems remain vulnerable to known exploits. Implementation of email authentication protocols including Sender Policy Framework (SPF), DomainKeys Identified Mail (DKIM), and Domain-based Message Authentication, Reporting and Conformance (DMARC) have proven effective at preventing email spoofing, reducing spam, and mitigating phishing attacks by verifying sender legitimacy~\cite{securware25}. Time to detection for compromised systems indicates both the effectiveness of monitoring capabilities and the duration of adversary access to sensitive data and systems.

\begin{figure}
    \centering
    \includegraphics[width=1.00\linewidth]{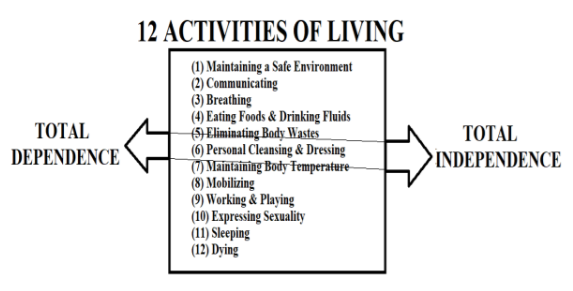}
    \caption{The 12 Activities of Daily Living ~\cite{mchugh}.}
    \label{fig:daily_living}
\end{figure}

Public health uses population-level indicators that transcend individual clinical measures. Beyond mortality and disease incidence, the field tracks broader indicators of population wellbeing. The concept of ``Activities of Daily Living'' (ADL) has been a cornerstone of nursing since the 1950s (see Figure~\ref{fig:daily_living}) ~\cite{adl}. ADLs comprise the essential activities individuals must perform to live independently: eating, bathing, dressing, managing personal hygiene, and so forth. These foundational tasks provide a framework for assessing health status and quality of life. Public health measures population health partly through the ability of individuals to perform these fundamental functions.

What are the analogous ``digital activities of daily living'' that increasingly define modern life? These would constitute meaningful indicators of the health of digital infrastructure. Online commerce has become essential for many; internet outages affecting e-commerce capability represent measurable impacts on digital daily living. GPS navigation enables reliable transportation. Digital communications including email underpin both professional and personal life. Access to government services, healthcare records, financial services, and educational resources increasingly occurs through digital channels. While not everyone relies on all these capabilities equally, disruptions to these functions represent measurable harms to populations who depend on them. Tracking the availability, reliability, and security of these digital activities would provide population-level health indicators for cyberspace, analogous to how public health tracks populations' ability to perform activities of daily living.

The key to public health's measurement capability is standardized reporting infrastructure. The U.S. Centers for Disease Control and Prevention (CDC) maintains surveillance systems that collect data from healthcare providers, laboratories, and public health departments using standardized case definitions. This infrastructure enables real-time monitoring of disease trends, early detection of outbreaks, and evaluation of intervention effectiveness. Cybersecurity lacks equivalent infrastructure. Incident reporting is voluntary, definitions vary across organizations, and no central body systematically collects, analyzes, and disseminates cybersecurity health metrics at a national scale.

Establishing these measurements, however, requires more than technical solutions or voluntary cooperation. As with public health, systematic measurement of cyber health requires coordinated infrastructure that transcends individual organizational capabilities. Public health's measurement systems did not emerge from market forces or voluntary cooperation among healthcare providers; they resulted from deliberate policy choices to create public infrastructure for population health surveillance. The same applies to cybersecurity. Without coordinated action to establish standardized definitions, reporting systems, and analysis capabilities, we will continue to lack the foundational measurements needed to assess progress or allocate resources effectively.

\section{Public Health and Cybersecurity as Public Goods}
\label{section:Argument for Government Involvement}

The preceding sections have documented cybersecurity's critical gaps: undefined populations, unmeasured outcomes, opaque transmission mechanisms, shocking knowledge deficits, and absent measurement infrastructure. These are not merely technical problems requiring better tools or methods. They are structural problems that arise from cybersecurity's economic characteristics.

Cybersecurity exhibits the classic properties of a \textit{public good} in economic terms. Public goods are non-excludable (one party's use does not prevent another's use) and non-rivalrous (consumption by one party does not diminish availability to others)~\cite{cowen2008public}. When an organization hardens its systems against sophisticated adversaries, it forces attackers to develop new capabilities, raising the bar for all subsequent attacks. This security improvement benefits others whether or not those others contributed to the cost. When researchers discover and disclose vulnerabilities, all users of affected software benefit. When organizations share threat intelligence, faster detection becomes possible across the ecosystem. These are classic positive externalities: costs borne by one actor create benefits that diffuse across many actors.

Similarly, cybersecurity information and defensive capabilities are largely non-rivalrous. When one organization implements email authentication protocols, the resulting reduction in spoofed emails benefits all recipients. When malware analysis techniques are published, this knowledge remains available to all defenders regardless of how many use it. In many cases, defensive value actually increases with broader adoption as network effects amplify protection.

This public good nature creates a fundamental market failure. Private organizations systematically underinvest in cybersecurity because they cannot fully capture the social value of their security improvements. Consider a financial institution investing heavily in defenses against nation-state adversaries. The institution bears the full cost, but benefits extend far beyond its balance sheet. By forcing adversaries to develop new techniques, it raises costs for all future attacks across its sector. By sharing threat intelligence with partners, it enables faster detection elsewhere. Yet the institution cannot monetize these broader benefits. Markets will under-provide such goods because rational actors invest only to the point where private benefit equals private cost, ignoring substantial positive externalities.

Public health operates under identical economics, which explains why vaccination programs receive public funding, why epidemiological surveillance is a government function, and why outbreak response involves public health departments rather than private healthcare providers alone. When individuals get vaccinated, they create herd immunity protecting unvaccinated community members. This classic positive externality means markets alone cannot optimize vaccination rates. The individual bears costs (time, potential side effects) while benefits accrue to the broader population. This is precisely why public health transformed from sporadic local efforts into a coordinated government function. The same economic logic applies to cybersecurity: relying exclusively on individual organizations or market-based solutions will systematically under-provide cybersecurity infrastructure, just as markets under-provide other public goods absent collective action.

This economic foundation explains the specific structural requirements that follow. A public health system, as described by the Institute of Medicine, is a complex network of public, private, and voluntary entities working together to protect population health~\cite{institute1988future}. The core functions are structured around three areas: (1) assessment, (2) policy development, and (3) assurance, accomplished through the 10 Essential Public Health Services~\cite{ephs}. Assessment involves monitoring population health status, collecting and analyzing data, and tracking vital statistics. Policy development encompasses informing, educating, and empowering people about health issues and behaviors. Assurance means improving and innovating through ongoing evaluation, research, and quality improvement.

We propose that the Federal Government has an essential role in establishing a comparable Cyber Public Health System. This follows from cybersecurity's public good characteristics, reinforced by several practical requirements:

\textbf{Scale and Coordination:} Cyber threats transcend individual organizations and sectors. The SolarWinds compromise affected thousands of organizations across government, technology, and defense through a single supply chain vulnerability. The Colonial Pipeline attack disrupted fuel distribution across the Eastern United States. These incidents demonstrate that effective cybersecurity requires coordination exceeding any individual entity's capacity. Only government possesses the authority and resources to coordinate at this scale.

\textbf{Critical Infrastructure Protection:} Government has constitutional responsibility to protect national security and public safety from sophisticated threats like cyber warfare and espionage. Critical infrastructure systems for energy, water, transportation, financial services, and healthcare are vulnerable to attacks causing widespread disruption. While private entities own much of this infrastructure, they lack resources and incentives to defend against nation-state adversaries whose capabilities far exceed typical criminal threats.

\textbf{Economic Stability:} Large-scale cyberattacks can disrupt businesses, cause substantial financial losses, and undermine public trust in the digital economy. Government must secure digital systems underpinning economic activity and social functioning. Market mechanisms alone cannot adequately address threats that could destabilize entire sectors or erode public confidence in digital transactions.

\textbf{Standardization and Data Infrastructure:} Effective cybersecurity requires standardized definitions, metrics, and reporting protocols. Individual organizations cannot unilaterally establish standards others will adopt. Government can set uniform standards for measurement, collection, reporting, and storage of cybersecurity data, enabling meaningful comparison, trend tracking, and evidence-based evaluation. Just as public health established standardized disease reporting enabling epidemiological analysis, cybersecurity needs standardized incident reporting for systematic improvement.

\textbf{Research Funding:} The Federal Government can provide sustained funding for cybersecurity research with broad public benefit but uncertain commercial returns. This mirrors public health research funding that has driven advances in disease prevention, treatment, and health systems organization.

\textbf{Coordinated Response:} When major incidents occur, coordinated response across organizations and sectors is essential. Government can establish institutional frameworks, communication channels, and resource allocation mechanisms for effective response, including stress tests, centralized monitoring, and economic incentives for improved cybersecurity.

\textbf{Information Asymmetries:} Attackers observe their techniques across many targets, learning what works. Defenders typically see only attacks against their own systems. Government-produced threat intelligence and government-coordinated sharing can help redress this imbalance but it requires trusted institutional frameworks and legal protections for sharing sensitive information~\cite{dykstra2023maximizing}.

\textbf{Long-Term Perspective:} Market pressures often drive short-term thinking when quarterly earnings cycles and competitive concerns lead organizations to underinvest in security measures with long-term benefits. The government can take a longer view, investing in infrastructure and capabilities taking years to demonstrate value but essential for sustained security.

The case for government involvement rests fundamentally on cybersecurity's public good nature, creating market failures that require collective action. This is reinforced by practical requirements for coordination, standardization, research, and response at scales beyond private sector capability. Just as public health transformed from sporadic local efforts into an evidence-based discipline through government leadership and investment, cybersecurity requires similar transformation. The question is not whether government should be involved, but how to structure involvement most effectively to achieve measurable improvements in national cyber health while preserving innovation and appropriate private sector roles.

\section{Summary}
\label{section:Summary}

In summary, this paper proposes the establishment of a Cyber Public Health System as a coordinated federal infrastructure for improving national cybersecurity. Our argument rests on three foundations. First, public health demonstrates that systematic data collection, evidence-based intervention, and coordinated response can achieve remarkable measurable improvements in population-level outcomes. Second, cybersecurity faces analogous challenges requiring similar systematic approaches: undefined populations, unmeasured outcomes, opaque transmission mechanisms, and fragmented defensive efforts. Third, cybersecurity shares public health's economic characteristics as a public good, creating market failures that necessitate government coordination rather than relying solely on private sector solutions.

While cybersecurity is often treated as a siloed IT problem, it is an enterprise-wide risk requiring a high-level department ensuring unified, multi-pronged approaches involving public health, law enforcement, and military agencies, aligning stakeholders across all sectors.

\section*{Acknowledgments}
\label{section:Acknowledgments}
We would like to thank 
L. Jean Camp, 
Yi Ting Chua, 
Brian LaMacchia, 
Daniel Lopresti, and
Adam Shostack 
for contributions to previous presentations of this material.

\bibliographystyle{IEEEtran}
\bibliography{bibliography}

@article{shostack2025lessons,
  title={Lessons for Cybersecurity from the American Public Health System},
  author={Shostack, Adam and Camp, L Jean and Chua, Yi Ting and Dykstra, Josiah and LaMacchia, Brian and Lopresti, Daniel},
  journal={2024-2025 CRA Quadrennial Paper},
  year={2025}, 
  url={https://cra.org/wp-content/uploads/2025/01/2024-2025-CRA-Quad-Paper_-Lessons-for-Cybersecurity-from-the-American-Public-Health-System.pdf}}

@techreport{Shostak_2024, 
title={
Inaugural Workshop on Cyber Public Health}, institution={CyberGreen Institute}, 
author={Shostack, Adam}, 
month={June}, 
number={24-01}, 
year={2024},
url={https://cybergreen.net/workshop-report-24-01-inaugural-workshop-on-cyber-public-health/}}

@inproceedings{dykstra2024handling,
  title={Handling Pandemic-Scale Cyber Threats: Lessons from COVID-19},
  author={Dykstra, Josiah and Shostack, Adam},
  booktitle={Proceedings of the New Security Paradigms Workshop (NSPW)},
  year={2024}}

@inproceedings{dykstra2023position,
  title={Position Paper: Evaluating Analogies and Applying Public Health Models for Cybersecurity},
  author={Dykstra, Josiah and Saydjari, O Sami and Met, Jamie and Hough, Douglas},
  booktitle={Proceedings of the 2024 Workshop on Cybersecurity in Healthcare (HeathSec)},
  year={2024}}

@inbook{spafmythbook,
  title={Cybersecurity Myths and Misconceptions: Avoiding the Hazards and Pitfalls that Derail Us},
  author={Spafford, Eugene and Metcalf, Leigh and Dykstra, Josiah},
  publisher={Addison-Wesley Professional},
year={2023}}

@book{topographyofcholera,
    author = {{S. V. Grevsmühl}},
title = {{Cartography: John Snow and the Topography of Cholera, within Cultures of Contagion (B. Delaurenti and T.L. Roux (editors)}},
 publisher = {{MIT Press, pp 55-63}},
    year = {2021}}

@article{cdcpaper,
  title={Achievements in Public Health, 1900-1999: Control of Infectious Diseases},
  author={{U. S. Centers for Disease Control (CDC)}},
  journal={Morbidity and Mortality Weekly Report (MMWR)},
  volume={48(29)},  
   year={1999},
  url={https://www.cdc.gov/mmwr/preview/mmwrhtml/mm4829a1.htm#:~:text=Disease%20control%20resulted%20from%20improvements%20in%20sanitation,foundation%20for%20today's%20disease%20surveillance%20and%20control}}

@article{cdcpaper2,
  title={Status Report on the Childhood Immunization Initiative: Reported Cases of Selected Vaccine-Preventable Diseases--United States},
  author={{U. S. Centers for Disease Control (CDC)}},
  journal={Morbidity and Mortality Weekly Report (MMWR)},
  volume={46},
  pages={665-671},
  year={1996}}

@article{owid-vaccines-children-saved,
    author = {Ritchie, Hannah},
    title = {Vaccines have saved 150 million children over the last 50 years},
    journal = {Our World in Data},
    year = {2024},
    url = {https://ourworldindata.org/vaccines-children-saved}}

@book{publichealthhistory,
    author = {{Institute of Medicine}},
    title = {The Future of Public Health},
    publisher = {The National Academies Press},
    year = {1988},
    url = {https://nap.nationalacademies.org/catalog/1091/the-future-of-public-health}}

@book{publichealthhistory2,
    author = {{Institute of Medicine}},
    title = {Emerging infections: microbial threats to health in the United States},
    publisher = {The National Academies Press},
    year = {1994},
    url = {https://wwwnc.cdc.gov/eid/pdfs/lederburg-report-2008.pdf}}

@article{johnsnowpaper,
    author = {T H Tulchinsky},
    title = {John Snow, Cholera, the Broad Street Pump; Waterborne Diseases Then and Now},
    journal = {Case Studies in Public Health},
    year = {2018},
    url = {https://pmc.ncbi.nlm.nih.gov/articles/PMC7150208/#fr0010}}

@book{adl,
    author = {{N. Roper, W.W. Logan, and A. J. Tierney}},
title = {The Roper-Logan-Tierney Model of Nursing},
 publisher = {Elsevier-Health Sciences Division},
    year = {2020}}

@article{mchugh,
    author = {{J. McHugh and W. Yurcik}},
    title = {Position Paper: Personal
Experience in the Technology Opportunities and Associated Risks
of Healthcare Challenges in a Continuing Care Retirement
Community (CCRC)},
    journal = {Proceedings of the ACM Cybersecurity in
Healthcare Workshop (HealthSec)},
    year = {2024},
url= {https://doi.org/10.1145/3689942.3694752}}

@article{securware25,
    author = {{ W. Yurcik, S. North, R. O'Kane, O.S. Saydjari, F.R. Miranda, R.S. Avelino,  and G. Pluta}},
    title = {Measurability: Toward Integrating Metrics into Ratings for Scalable Proactive Cybersecurity Management},
    journal = {{IARIA Nineteenth International Conference on Emerging Security Information, Systems and Technologies (SECURWARE)}},
    year = {2025}}

@misc{ephs,
     author = {{U.S. Centers for Disease Control and Prevention}},
     title = {10 Essential Public Health Services},
        url = {https://www.cdc.gov/public-health-gateway/php/about/index.html}}

@techreport{cdc2023life, 
title={
United States Life Tables, 2021}, institution={U.S. Department of Health and Human Services}, 
author={Elizabeth Arias and Jiaquan Xu and Kenneth Kochanek}, 
month={November}, 
volume={72},
number={12}, 
year={2023},
url={https://www.cdc.gov/nchs/data/nvsr/nvsr72/nvsr72-12.pdf} 
}

@article{cowen2008public,
  title={Public goods},
  author={Cowen, Tyler},
  journal={The concise encyclopedia of economics},
  pages={197--199},
  year={2008}
}

@book{institute1988future,
  title={The future of public health},
  author={Institute of Medicine},
  year={1988},
  publisher={National Academies Press},
  address={Washington, DC}
}

@article{chen1999health,
  title={Health as a global public good},
  author={Chen, Lincoln C and Evans, Tim G and Cash, Richard A and others},
  journal={Global public goods},
  volume={1},
  pages={284--304},
  year={1999},
  publisher={Oxford University Press New York}
}

@article{anderson2006economics,
  title={The economics of information security},
  author={Anderson, Ross and Moore, Tyler},
  journal={science},
  volume={314},
  number={5799},
  pages={610--613},
  year={2006},
  publisher={American Association for the Advancement of Science}
}

@phdthesis{naghizadeh2016provision,
  title={On the Provision of Public Goods on Networks: Incentives, Exit Equilibrium, and Applications to Cyber},
  author={Naghizadeh Ardabili, Parinaz},
  year={2016},
  school={University of Michigan}
}

@inproceedings{anderson2001information,
  title={Why information security is hard-an economic perspective},
  author={Anderson, Ross},
  booktitle={Seventeenth annual computer security applications conference},
  pages={358--365},
  year={2001},
  organization={IEEE}
}

@incollection{varian2004system,
  title={System reliability and free riding},
  author={Varian, Hal},
  booktitle={Economics of information security},
  pages={1--15},
  year={2004},
  publisher={Springer}
}

@techreport{cybergreen2022public,
 title={{Public Health \& Cyber Public Health}},
 number={22-01},
 howpublished={\url{https://cybergreen.net/technical-report-22-01/}}, 
 author={Shostack, Adam}, 
 institution={CyberGreen Institute},
 year={2022}, 
 month={3}
}

@article{dykstra2023maximizing,
  title={Maximizing the benefits from sharing cyber threat intelligence by government agencies and departments},
  author={Dykstra, Josiah and Gordon, Lawrence A and Loeb, Martin P and Zhou, Lei},
  journal={Journal of Cybersecurity},
  volume={9},
  number={1},
  pages={tyad003},
  year={2023},
  publisher={Oxford University Press}
}

\end{document}